# Stability and Hydrolyzation of Metal Organic Frameworks with Paddle-Wheel SBUs upon Hydration

Kui Tan,† Nour Nijem,† Pieremanuele Canepa,‡ Qihan Gong,§ Jing Li,§ Timo Thonhauser‡ and Yves J Chabal*†

†Department of Materials Science & Engineering, University of Texas at Dallas, Richardson, Texas 75080

‡Department of Physics, Wake Forest University, Wake Forest Road, Winston-Salem, North Carolina, 27109

§Department of Physics and Astronomy and Department of Chemistry and Chemical Biology, Rutgers University, 610 Taylor Road, Piscataway, New Jersey 08854

Supporting Information

**ABSTRACT:** Instability of most prototypical metal organic frameworks (MOFs) in the presence of moisture is always a limitation for industrial scale development. In this work, we examine the dissociation mechanism of microporous paddle wheel frameworks M(bdc)(ted)$_{0.5}$ [M=Cu, Zn, Ni, Co; bdc= 1,4-benzenedicarboxylate; ted= triethylenediamine] in controlled humidity environments. Combined *in-situ* IR spectroscopy, Raman, and Powder x-ray diffraction measurements show that the stability and modification of isostructual M(bdc)(ted)$_{0.5}$ compounds upon exposure to water vapor critically depend on the central metal ion. A hydrolysis reaction of water molecules with Cu-O-C is observed in the case of Cu(bdc)(ted)$_{0.5}$. Displacement reactions of ted linkers by water molecules are identified with Zn(bdc)(ted)$_{0.5}$ and Co(bdc)(ted)$_{0.5}$. In contrast, Ni(bdc)(ted)$_{0.5}$ is less susceptible to reaction with water vapors than the other three compounds. In addition, the condensation of water vapors into the framework is necessary to initiate the dissociation reaction. These findings, supported by supported by first principles theoretical van der Waals density functional (vdW-DF) calculations of overall reaction enthalpies, provide the necessary information for determining operation conditions of this class of MOFs with paddle wheel secondary building units and guidance for developing more robust units.

## 1. Introduction

Metal organic frameworks (MOFs) are a family of nanoporous materials that are attracting great interest as a potential material for gas storage and separation.[1,2] These relatively new porous materials result from the reaction between organic and inorganic species that can assemble into one-, two- or three-dimensional structures.[3] The two subunits within the structure contains inorganic coupling parts, also referred to secondary building units (SBUs), and organic linkers, such as dicarboxylate or other organic rings.[4] A great variety of cations can be incorporated into the framework and a wide choice of functionalized organic linkers provide many possibilities to design the structures and tune the properties of the final porous materials.[5,6]

The high surface areas, porosity and tunable structures make this type of materials very promising for a variety of applications including gas storage, separation, sensing, catalysis and drug delivery.[3,7-16] Efforts have been done to explore MOFs as a sorbent for H$_2$ storage and CO$_2$ capture materials. H$_2$ storage in MOFs materials is usually achieved thorough fast physical adsorption onto the surface of pores with a large adsorption capacity.[16] MOFs are also viewed as an ideal platform for next generation CO$_2$ capture materials owing to its large adsorption capacity and chemical tunability of the interaction between CO$_2$ and adsorbents.[15] Despite these interesting applications, these studies have pointed out some issues that must be evaluated and addressed before MOFs is used in real-world systems. One of the major concern for using MOFs in gas storage and separation is their chemical stability under humid condition.[15,16] The stability of MOFs under water vapor atmosphere has so far been mainly evaluated by X-ray Diffraction.[17-19] Some widely investigated MOFs including MOF-5, MOF-177, though exhibiting excellent performance in H$_2$ and CO$_2$ storage in porous materials, are not stable in the presence of small amount of water.[20-22] Greathouse and Allendorf used empirical force fields and molecular dynamics to simulate the interaction of water with MOF-5.[21] They predicted that the low stability of MOF-5 in water is due to the weak interaction between Zn and O atoms in the structure. This is consistent with the experimental results of Huang, who proposed that water can strongly bind to the framework, leading to hydrolysis and the formation of terephthalic acid.[20] In other studies, Low modeled the reaction of water with metal organic frameworks from first principles and determined hydrolysis products.[18]

While some theoretical work has been done to understand the interaction and possible reaction of water molecules with MOFs, there are no experimental studies to validate these theoretical model and give the detailed information of structural decomposition of metal organic framework.[23] Precise characterization of reaction of water molecules with frameworks building units such as breaking and reforming bonds is extremely important to obtain insight into the mechanism of MOFs dissociation process in humid environments.

This work focuses on water interaction with one prototypical metal organic framework, M(bdc)(ted)$_{0.5}$ [M=Cu, Zn, Ni, Co, H$_2$bdc = 1,4-benzenedicarboxylate; ted = triethylenediamine], that contains SBUs of two 5-coordinate copper cations bridged in a paddle wheel-type configuration.[24-31] (See Fig. 1 and Fig. S1). M(bdc)(ted)$_{0.5}$ has been shown to be an excellent absorbent for a variety of gases (H$_2$, CO$_2$, CH$_4$ and other hydrocarbons).[29-31] The paddle wheel binuclear metal clusters within M(bdc)(ted)$_{0.5}$ can be found in a large number of reported nanoframeworks materials with high microporosity, thermal stability, good sorption properties and other advanced functional properties such as magnetic property.[32-37] Using new pillaring or mixed ligands, it is possible to prepare a variety of functionalized and flexibly tuned 3D architectures by altering the nature of carboxylate and pillaring ligands.[31, 38, 39] We combine (i) *in-situ* infrared (IR) spectroscopy, (ii) *ex-situ* Raman spectroscopy, and (iii) powder X-ray diffraction to investigate the reaction of water molecules with the prototypical frameworks M(bdc)(ted)$_{0.5}$ with paddle-wheel type secondary building units (SBUs). IR and Raman spectroscopy techniques have proved to be useful in studying weak interactions of small molecules in MOFs such as hydrogen and CO$_2$ in MOFs.[40-44] and in providing information about water bonding, as previously demonstrated for characterizing adsorbed water in MOF materials.[45, 46] Raman spectroscopy provides complementary information to IR, and enables the characterization of metal oxide vibrational modes of the MOF that occur in the low frequency region (50-600 cm$^{-1}$).[47]

In this work, IR, Raman and X-ray diffraction measurements are combined with first principles theoretical van der Waals density functional (vdW-DF) calculations. We find that the nature of the central metal ion in M(bdc)(ted)$_{0.5}$ critically affects the stability and decomposition pathways in humid environments.

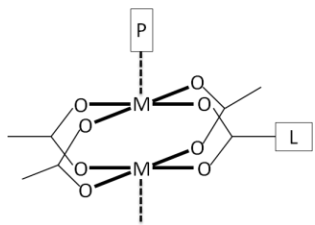

**Figure 1.** Schematic illustration of coordination geometry of paddle wheel building units: M = Metal ion, L = Bicarboxylate linker and P = N-containing bidentate pillar linker.

## 2. Experimental and Theoretical Methods

Synthesis.

Cu(bdc)(ted)$_{0.5}$.

A mixture of copper (II) nitrate trihydrate (0.039 g), H$_2$bdc(0.068 g), ted (0.048 g) and 15 ml of DMF was transferred into a 20 mL vessel. The vessel was then sealed and heated to 120 °C overnight. A blue color crystalline powder was obtained. After filtering and washing with 20 mL DMF, the product was collected. Then the sample was heated at 120 °C again under a flow of dry N$_2$ for one day to remove the guest DMF molecules.

Zn(bdc)(ted)$_{0.5}$.

A mixture of zinc (II) nitrate hexahydrate (0.245 g, 0.82 mmol), H$_2$bdc (0.136 g, 0.97 mmol), ted (0.073 g, 0.65 mmol) and 15 mL of DMF were transferred to Teflon-lined autoclave and heated at 120 °C for 2 days. Cubic colorless crystals of Zn(bdc)(ted)$_{0.5}$ were isolated by filtering and washed three times with 10 mL of DMF. Then the sample was heated at 120 °C again under a flow of dry N$_2$ for one day to remove the guest DMF molecules.

Ni(bdc)(ted)$_{0.5}$.

A mixture of nickle (II) chloride hexahydrate (0.107 g, 0.45 mmol), H$_2$bdc (0.060 g, 0.36 mmol), ted (0.033 g, 0.29 mmol) and 15 ml of DMF were transferred to Teflon-lined autoclave and heated at 120 °C for 2 days. Green crystalline powder of Zn(bdc)(ted)$_{0.5}$ was isolated by filtering and washed three times with 10 mL of DMF. Then the sample was heated at 120 °C again under a flow of dry N$_2$ for one day to remove the guest DMF molecules.

Co(bdc)(ted)$_{0.5}$.

A mixture of cobalt (II) nitrate hexahydrate (0.13 g, 0.45 mmol), H$_2$bdc (0.060 g, 0.36 mmol), ted (0.033 g, 0.29 mmol) and 15 mL of DMF were transferred to Teflon-lined autoclave and heated at 120 °C for 2 days. Purple crystalline powder of Zn(bdc)(ted)$_{0.5}$ was isolated by filtering and washed three times with 10 mL of DMF. Then the sample was heated at 120°C again under a flow of dry N$_2$ for one day to remove the guest DMF molecules.

Infrared spectroscopy:

A powder of M(bdc)(ted)$_{0.5}$ (~2 mg) was pressed onto a KBr support and placed into a high pressure high temperature cell purchased from Specac at the focal point of the sample compartment of infrared spectrometer (Nicolet 6700, Thermo Scientific) equipped with a liquid N$_2$–cooled MCT-B detector. The cell was connected to a vacuum line for evacuation. To avoid interference between the infrared absorption bands of trace amount of unreacted H$_2$bdc molecules and of H$_2$O, D$_2$O vapor was used in the water vapor exposure experiments aimed at observing hydrolysis products of BDC ligands. All spectra were recorded in transmission between 400 and 4000 cm$^{-1}$ (4 cm$^{-1}$ spectral resolution). A gas phase reference was taken at each pressure with a pure KBr pellet as reference for subtraction.

Raman Spectroscopy.

The Raman measurements were collected using a Nicolet Almega XR Dispersive Raman spectrometer from Thermofisher. A 532 nm solid state laser was used for excitation. The output power was reduced to 10% (1.23 mW) and the acquisition time varied from 5 to 10 min to avoid sample decomposition. The spectra were obtained from 50 to 2000 cm$^{-1}$ with a resolution of 0.9642 cm$^{-1}$. The spectrometer was equipped with a 50 objective microscope. A larger quantity of M(bdc)(ted)$_{0.5}$ powder (~8 mg) exposed to similar conditions to the IR experiments was used for ex situ Raman and XRD measurements.

X-ray powder diffraction.

Out of plane X-ray powder diffraction data were recorded in the 2 theta mode from 5° to 50° on Rigaku Ultima III diffractometer (Cu Kα radiation, X-ray wavelength of 1.5418 Å, operating at 40 keV with a cathode current of 44 Ma).

vdW-DF calculations.

Calculations were performed using the vdW-DF functional as available in the *PWscf* code (a Quantum ESPRESSO package).[48] The vdW-DF[49-51] functional has been recently shown as a very promising tool for describing van der Walls forces acting in the adsorption of some non-polar molecules in MOF.[52] Ultrasoft pseudopotentials together with plane-waves, with an optimized cutoff of 35 Ry, were used to describe the charge density of the M(bdc)(ted)$_{0.5}$ MOF. An accurate sampling of potential energy surface was ensured by imposing tight thresholds on both the total energy, $5*10^{-12}$ Ry, and forces, $5*10^{-4}$ Ry Bohr$^{-1}$. The total energy was sampled on 2 2 2 k-point grid according to the Monkost-Pack scheme.[53] Our analysis is only restricted to the Ni and Zn(bdc)ted$_{0.5}$ structures. Experimentally the Zn(bdc)(ted)$_{0.5}$ structure is affected by some proton statistical disorder (previously solved by Kong *et al.*[52]) resulting in a tetragonal cell with *a* = *b* = 10.93 and *c* = 9.61 Å. The Ni(bdc)(ted)$_{0.5}$ was obtained after full relaxation of the Zn(bdc)(ted)$_{0.5}$ where Zn atoms were replaced by Ni atoms. The optimized lattice constants for Ni(bdc)(ted)$_{0.5}$ are *a* =, *b* = 11.15 and *c* = 9.53 Å. Our initial investigation on the Ni(bdc)ted$_{0.5}$ structure shows that the magnetic moment on the Ni atoms (2 per cell) is 1.76 $\mu_b$ Ni$^{-1}$ and remains almost totally localized on the Ni species

3. Results

3.1 IR Spectroscopy

The assignment of the MOFs vibrational modes is a necessary step toward determining the structural changes occurring upon gas uptake from the vibrational spectra. The structure of Cu(bdc)(ted)$_{0.5}$ contains two organic linkers 1,4-benzenedicarboxylic acid, triethylenediamine and metal oxide clusters. The IR spectrum of activated Cu(bdc)(ted)$_{0.5}$ is dominated by the bands associated with the organic parts of the MOF. The infrared spectra of the bdc acid and its related compound have been studied in-depth in the past few years.[54-56] The fundamental vibrations of ted molecules in the gas, solution and solid phases have also previously been assigned.[57] These studies, together with spectra of pure organic ligands of H$_2$bdc acids and ted linkers, make it possible to derive a detailed assignment of the MOF vibrational modes. In the paddlewheel structure, the carboxylate ion COO$^-$ is coordinated to two Cu atoms as a bridging bidentate ligand in a syn-syn configuration as shown in Fig. 1. The –COO$^-$ group is characterized by in phase or out of phase vibrations of the two equivalent C-O bonds, leading to the symmetric and antisymmetric stretch modes. Previous work has suggested that the splitting Δ between these two ν(COO) frequencies is related to the nature of the carboxylate coordination.[58, 59] For example, carboxylate with bridging bidentate configuration gives values in the range of 160-200 cm$^{-1}$ similar to vibrations of ionic species.[58] The intensity of the asymmetric mode is usually larger than that of symmetric stretching mode.[60] Large splittings (>200 cm$^{-1}$) are usually associated with monodentate carboxylate due to a change from equivalent to inequivalent carbon-oxygen bonds.[58] In bidentate bridged carboxylate-Cu(II) complexes, the ν$_{as}$(COO) species is usually characterized by a wavenumber in the 1600-1630 cm$^{-1}$region.[61, 62] On the basis of all this previous work, we assign the band at 1622 cm$^{-1}$, overlapped with an intense band at 1575 cm$^{-1}$, to out of phase (IR active) motion of asymmetric stretch ν$_{as}$(COO). The intense band at 1575 cm$^{-1}$, as well as those at 1507 cm$^{-1}$, 1152 cm$^{-1}$ and 1017 cm$^{-1}$ are all associated with the phenyl modes, based on the comparison with pure bdc ligands and reference assignments.[54-56] Considering the relationship between the separation Δ of the two ν(COO) frequencies and the nature of the carboxylate coordination, we infer that, since the ν$_{antisym}$(COO) is at 1622 cm$^{-1}$, the symmetric stretch mode of ν$_{sym}$(COO) should be higher than 1420 cm$^{-1}$ and the intensity of $\nu_{sym}$(COO) weaker than the band at 1622 cm$^{-1}$. Based on this consideration, we assign the shoulder at 1430 cm$^{-1}$ rather than the strong peak at 1391 cm$^{-1}$ to $\nu_{sym}$(COO).

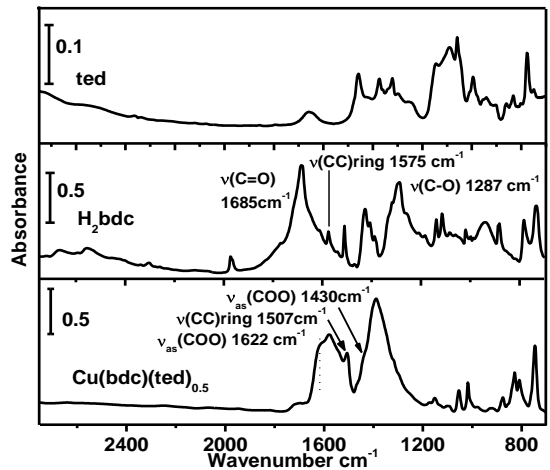

**Figure 2**. IR adsorption spectra of ted, H$_2$bdc, and activated Cu(bdc)(ted)$_{0.5}$ in vacuum referenced to KBr in vacuum.

Table 1. Selected IR, Raman Frequencies ( cm$^{-1}$) and their assignment for Cu(bdc)(ted)$_{0.5}$.[a]

| IR | Raman | Assignment |
| --- | --- | --- |
| 1622, | 1535 | $\nu_{as}$(COO) |
| 1575, 1507 | | 19a, 19b, 18b and 18a |
| 1152, 1017 | | |
| | 1617 | 8a CC stretching |
| 1430 | 1442, 1430 | $\nu_s$(COO) |
| 1054 | | $\nu_{as}$(NC$_3$) |
| 874 | | $\gamma$(CH)$_{oop}$ |
| 828 | | $\rho$(CH$_2$) |
| 810 | | $\delta$(COO)op |
| | 727 | $\delta$(COO)ip |
| 744 | | 12 ring deformation |
| | 398, 316 | $\nu$(Cu-O) |

[a]The numbers 19a, 19b, 18b, 18a, 8a and 12 represents Wilson notation to describe the phenyl vibrational modes: 19a and 19b contain the mixture of the $\nu$(C=C) and CH bending motions; 18a and 18b are classified as the mode having CH bending character. 8a mode is assigned to CC stretching under C$_i$ symmetry. 12 is attributed to the benzene trigonal ring deformation.

This intense band at 1391 cm$^{-1}$ also appears in spectrum of the Cu-btc sample (btc=1,3,5-benzenetricarboxylate) and Ni-based MOFs containing paddle-wheel type inorganic building units, and is therefore attributed to the benzene ring mode.[37, 62, 63] Since it is not directly connected with the central metal ions, it is the least affected by substitution of central metal ions. The presence of this bands at the same frequency (1391 cm$^{-1}$) in the other three M(bdc)(ted)$_{0.5}$ series [M=Zn, Ni, Co] indicates that it is independent of metal centers like the band at 1507 cm$^{-1}$. The band at 810 cm$^{-1}$ is assigned to the $\delta$(COO)oop (IR active), while the Raman active band $\delta$(COO)ip is at 727 cm$^{-1}$. Other assignments are listed in Table 1. By applying the same reasoning, we assign the other M(bdc)(ted)$_{0.5}$ series, as shown in Fig. S3 (Supporting Information).

To select the experimental conditions, we note that Liang's H$_2$O vapor adsorption isotherm and X-ray powder pattern studies showed that the structures of two isostructural MOFs M(bdc)(ted)$_{0.5}$ (M = Zn, Ni) remain stable under 30% relative humidity at 25 °C, yet collapse under 60% relative humidity.[64]

For our work we therefore chose two pressures: 6 Torr, the common relative humidity 30% in atmosphere and 9.5 Torr D$_2$O vapor, relative humidity 48% at room temperature of 24 °C.

After introduction of 6 Torr D$_2$O at room temperature, a broad band associated with the stretching modes of adsorbed D$_2$O appears in the region from 2700 to 1800 cm$^{-1}$ in all four

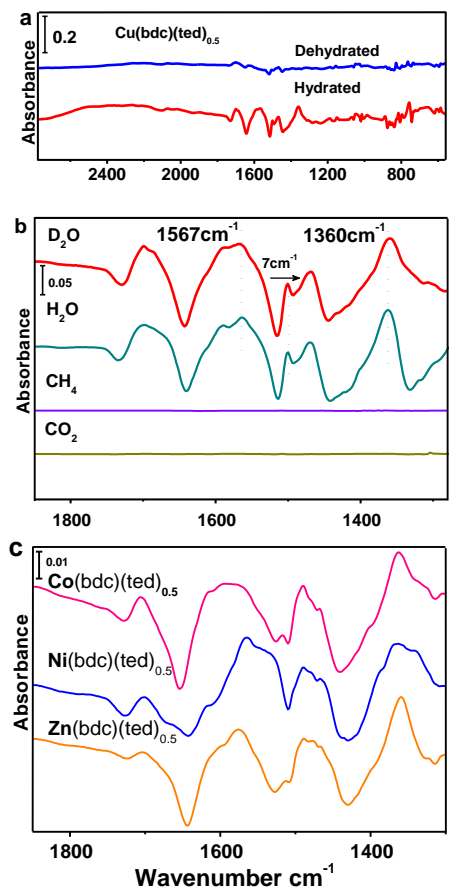

**Figure 3.** (a): IR absorption spectra of Cu(bdc)(ted)$_{0.5}$ exposure to 6 Torr D$_2$O vapor (red spectrum), dehydrated Cu(bdc)(ted)$_{0.5}$ after evacuation at 413 K (blue spectrum). Both spectra are referenced to activated Cu(bdc)(ted)$_{0.5}$. (b): The middle frequency range of Cu(bdc)(ted)$_{0.5}$ exposure to 6 Torr D$_2$O vapor, H$_2$O vapor, CH$_4$, and CO$_2$. (c): IR absorption spectra of Zn(bdc)(ted)$_{0.5}$, Ni(bdc)(ted)$_{0.5}$, and Co(bdc)(ted)$_{0.5}$ exposure to 6 Torr D$_2$O.

compounds (See Fig. 3a and Fig. S15). Furthermore, the IR absorption spectrum of Cu(bdc)(ted)$_{0.5}$ is strongly affected after D$_2$O exposure in the middle frequency range from 1800 to 1300 cm$^{-1}$, away from the D$_2$O bending mode region (around 1200 cm$^{-1}$). For this MOF [Cu(bdc)(ted)$_{0.5}$], a red shift of the $\nu$(C=O) mode from 1730 to 1700 cm$^{-1}$ is due to deuterium bonding of the C=O bond with the incoming D$_2$O molecule. The red shift of 7 cm$^{-1}$ of the mode at 1507 cm$^{-1}$ (Fig. 3b) is related to perturbation of the benzene ring stretching band 19b of Cu(bdc)(ted)$_{0.5}$. The two new features at 1360 cm$^{-1}$ and 1567 cm$^{-1}$ arising upon introduction of D$_2$O molecules are attributed to $\nu_{as}$(COO) and $\nu_s$(COO), initially at ~1600-1640 cm$^{-1}$ and

~1410-1430 cm$^{-1}$, as discussed in next section. To distinguish and identify the role of water vapor in inducing these spectral changes, nonpolar gases (CH$_4$, CO$_2$) exposures were also performed under the same conditions. As shown in Fig. 3b, little change results from non-polar CO$_2$ and CH$_4$ molecules. In contrast, the same large perturbations for Cu(bdc)(ted)$_{0.5}$ are noted upon H$_2$O exposure upon D$_2$O exposure. As shown Fig. 3c, the same perturbation bands are identified in the frequency region of 1800 to 1350 cm$^{-1}$ in the other three M(bdc)(ted)$_{0.5}$ [M=Zn, Ni, Co] compounds. Further experiments of the pressure dependence of D$_2$O adsorption into M(bdc)(ted)$_{0.5}$ (see Fig. S15) also confirm that these intensity variations are associated with incorporation of D$_2$O molecules into the frameworks. This conclusion is confirmed by evacuating the sample at 413 K overnight after D$_2$O gas loading at 6 Torr, and noting that both D$_2$O-related absorption and all spectral perturbations are nearly removed.

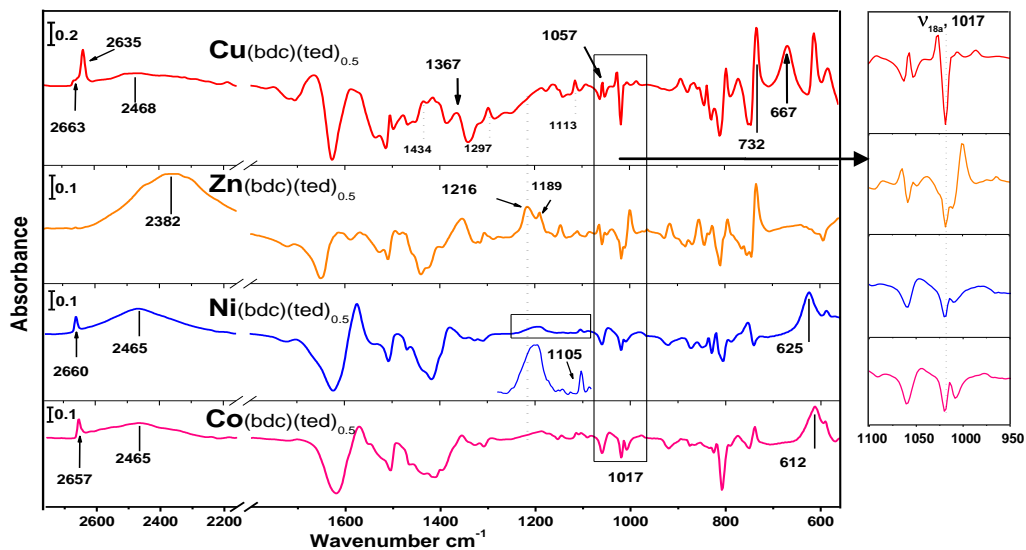

**Figure 4**. IR adsorption spectra of hydrated M(bdc)(ted)$_{0.5}$, referenced to activated MOF in vacuum after introduction of 9.5 Torr D$_2$O vapor and evacuation of gas phase D$_2$O vapor. Color scheme: red, Cu(bdc)(ted)$_{0.5}$; black, Zn(bdc)(ted)$_{0.5}$, blue, Ni(bdc)(ted)$_{0.5}$; pink, Co(bdc)(ted)$_{0.5}$.

Indeed, after exposing the activated sample to 9.5 Torr D$_2$O, the spectra of all four compounds exhibit distinctive features in the frequency range of 2700 - 2100 cm$^{-1}$ (Fig. 4). Sharp components can be seen in Cu(bdc)(ted)$_{0.5}$, Ni(bdc)(ted)$_{0.5}$, and Co(bdc)(ted)$_{0.5}$ at the frequencies of 2635, 2660, and 2657 cm$^{-1}$. In the case of Cu(bdc)(ted)$_{0.5}$, a small shoulder on a sharp band is observable at 2663 cm$^{-1}$ and attributed to O-D groups as described below.

The lineshape and position of the broad ν(O-D) absorption bands contain useful information to understand the interaction of adsorbed D$_2$O with the MOF framework. In Cu(bdc)(ted)$_{0.5}$, Ni(bdc)(ted)$_{0.5}$, and Co(bdc)(ted)$_{0.5}$, the band is centered between 2460 to 2480 cm$^{-1}$. However, in the case of Zn(bdc)(ted)$_{0.5}$, it is centered around 2380 cm$^{-1}$, a distinctively lower frequency. In the middle frequency range of 1800 to 950 cm$^{-1}$, the spectrum of Cu(bdc)(ted)$_{0.5}$ displays additional adsorption bands, compared to the other three MOFs. These bands are labeled by wavenumbers in Fig. 4. Similar to the stretching mode of broad ν(O-D) bands, the bending mode of D$_2$O in Zn(bdc)(ted)$_{0.5}$ is also distinctively different from that in the other three compounds: a peak develops at higher frequency (1216 cm$^{-1}$). The ν$_{18a}$ mode of the bdc ligand situated at 1017 cm$^{-1}$ in four compounds responds to hydration by different ways: it blue shifts to 1027 cm$^{-1}$ in the case of Cu(bdc)(ted)$_{0.5}$ but red shifts to 1013 cm$^{-1}$ in Ni(bdc)(ted)$_{0.5}$ and Co(bdc)(ted)$_{0.5}$; in the case of Zn(bdc)(ted)$_{0.5}$, it undergoes a larger red shift to 1001 cm$^{-1}$. In fact, this band is sensitive to the MOF structural changes and has been used before to monitor the transformation of flexible frameworks MIL-53 [CrIII(OH)(OOC–C$_6$H$_4$–COO)] during CO$_2$ adsorption and dehydration processes.[46, 65] In the low frequency range below 800 cm$^{-1}$, a broad band is observed at 667 cm$^{-1}$ in the Cu(bdc)(ted)$_{0.5}$ but absent in other three compounds. In Ni(bdc)(ted)$_{0.5}$ and Co(bdc)(ted)$_{0.5}$, new bands appear at 626 cm$^{-1}$ and 612 cm$^{-1}$ only after exposure to 9.5 Torr D$_2$O vapor (i.e. they are absent in the low vapor exposure regime).

In the desorption process, the removal of trapped D$_2$O requires a higher annealing temperature for Zn(bdc)(ted)$_{0.5}$ than for the other MOFs of the series (Cu, Ni, Co). Fig. 5 shows a slow release of bound D$_2$O molecules in Zn(bdc)(ted)$_{0.5}$ up to 200 °C. For the other MOF compounds, the D$_2$O molecules are completely removed at 100 °C under vacuum.

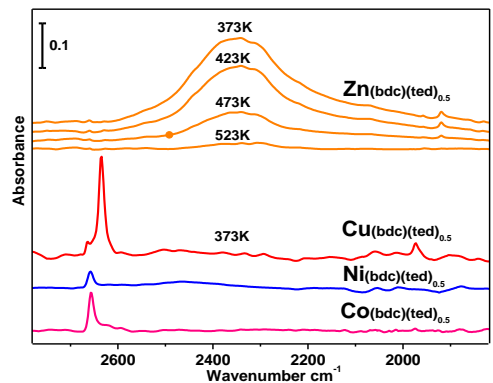

**Figure 5**. IR spectra of hydrated M(bdc)(ted)$_{0.5}$ [M=Zn, Cu, Ni, Co] after evacuation at elevated temperature, referenced to activated sample recorded at RT. Color scheme: Orange, Zn(bdc)(ted)$_{0.5}$ after outgassing for 10 h at 373 K, 423 K, 473 K, and 2 h at 523 K; Red, blue, and pink corresponding to Cu(bdc)(ted)$_{0.5}$, Ni(bdc)(ted)$_{0.5}$ and Co(bdc)(ted)$_{0.5}$ after outgassing for 10 h at 373 K.

### 3.2 Raman Spectroscopy

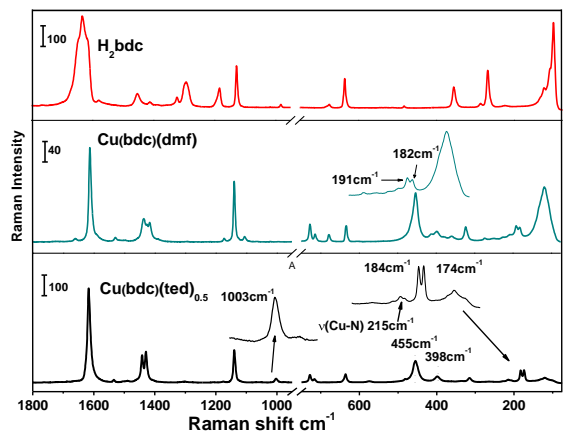

**Figure 6**. Raman spectra from top to bottom: H$_2$bdc acid, Cu(bdc)(dmf) and Cu(bdc)(ted)$_{0.5}$.

The vibrational assignments of M(bdc)(ted)$_{0.5}$ are best performed by comparing the Raman spectra of M(bdc)(ted)$_{0.5}$ to those of free H$_2$bdc molecules, ted ligands and the counterpoint M(bdc)(dmf) in Fig 6. For example, Cu(bdc)(dmf) was synthesized in the same conditions as Cu(bdc)(ted)$_{0.5}$ in the absence of ted ligands. The structure of Cu(bdc)(dmf), resembling the structure of Cu(bdc)(ted)$_{0.5}$, is formed of a three-dimensional network structure bridging the two-dimensional layers of porous copper(II) dicarboxylate with DMF solvent instead of pillar ligand of ted shown in middle part of Fig. 6.[66] In the high frequency range (1800 to 900 cm$^{-1}$), most of the Raman bands are due to vibrational modes of bdc linkers. The band at 1617 cm$^{-1}$ is assigned to the phenyl mode $\nu_{8a}$ [C=C stretching mode of bdc] that only appears in the Raman spectra because it has g parity under C$_i$ symmetry.[54] While the feature at 1535 cm$^{-1}$ is due to in-plane $\nu_{asym}$(COO), the bands at 1442 cm$^{-1}$ and 1430 cm$^{-1}$ can be assigned to $\nu_{sym}$(COO). The vibrational frequencies of the symmetric and antisymmetric $\nu$(COO) modes are different from the corresponding modes detected in IR measurements because they are in-phase (Raman-active) motions of asymmetric and symmetric CO$_2$ stretching, instead of out-of-phase (IR active).

In the low frequency region (750 to 50 cm$^{-1}$), Cu(bdc)(ted)$_{0.5}$ exhibits a more complex Raman spectrum than that of acid molecules because of the presence of vibrational modes of metal (Cu) oxide clusters. The vibrational features of metal oxide clusters can be assigned by comparing to Cu(bdc)(dmf), copper acetate monohydrate Cu(OAc)$_2$(H$_2$O)$_2$ [67], exhibiting the same [Cu$_2$C$_4$O$_8$] cage as Cu(bdc)(ted)$_{0.5}$, copper paddle-wheel-based MOFs Cu-btc [btc=1,3,5-benzenetricarboxylate],[47] and the M(bdc)(ted)$_{0.5}$ [M = Zn, Ni, Co] series in Fig. S4. Two bands at 455 cm$^{-1}$ and 398 cm$^{-1}$ are definitely attributed to Cu-O species and the 316 cm$^{-1}$ mode can be ascribed to $\nu$(Cu-O), which is also observed in Cu(bdc)(dmf) and Cu-btc compounds. A doublet appears in Cu(bdc)(ted)$_{0.5}$ at 184 and 174 cm$^{-1}$, in Cu(bdc)(dmf) at 191 and 182 cm$^{-1}$ as well as in Cu-btc at 193 and 177 cm$^{-1}$ but is absent in pure H$_2$bdc ligands. Prestipino ascribed these bands to a mode involving Cu-Cu stretching of the two Cu(II) ions of Cu$_2$[COO]$_4$ framework cage.[47] For the other three M(bdc)(ted)$_{0.5}$[M=Zn, Ni, Co] compound and Zn(bdc)(dmf), the doublet mode can be also readily identified around 190-160 cm$^{-1}$ in Fig.7 and Fig. S5. These doublets can be regarded as a unique feature of di-nuclear paddle wheel building units. The band at 215 cm$^{-1}$ can be tentatively assigned to $\nu$(Cu-N) bond.[67]

The full Raman spectra for another three compounds M(bdc)(ted)$_{0.5}$ [M = Zn, Ni, Co] are presented in the Supporting Information (See Fig. S4). Here what we are interested in are the two regions of 950 to 1200 cm$^{-1}$ and 100 to 500 cm$^{-1}$. In four compounds, we can observe a band around 1000 to 1050 cm$^{-1}$ in Fig. 7 for all activated MOFs. This feature is found neither in Cu(bdc)(dmf), Zn(bdc)(dmf) structures nor in pure bdc ligands (See Fig. S5). It must therefore be associated with the ted ligands. Previous studies have assigned such a band to $\nu_4$ ($\nu$CC/$\omega$CH$_2$) that involves a substantial amount of C-C stretching vibration.[68-70] In triethylenediamine molecules, the two nitrogen lone pairs are connected by -CH$_2$-CH$_2$- chains and interact primarily via through bond coupling.[71, 72] The mixing of the lone pair molecule orbitals with C-C molecule orbitals makes the $\nu_4$ ($\nu$CC/$\omega$ CH$_2$) mode frequency extremely sensitive to the changes in the environment of the lone pairs. For the unprotonated species of ted, $\nu_4$ is found at 983 cm$^{-1}$.[68] Upon coordination to metal ions Cu$^{2+}$, Co$^{2+}$, Ni$^{2+}$, Zn$^{2+}$, this band shifts to 1003 cm$^{-1}$, 1009 cm$^{-1}$, 1012 cm$^{-1}$, and even 1018 cm$^{-1}$. The coordination of the lone pair to metal ions causes a redistribution of the electron density, releasing the through-bond coupling and resulting in a blue shift of $\nu_4$ in four compounds.

After exposing the activated sample to 9.5 Torr D$_2$O vapor, the Raman spectra of hydrated M(bdc)(ted)$_{0.5}$ [M = Zn, Ni, Co] show marked changes compared to that of pristine sample (Fig. 7 and Fig. S6). For Cu(bdc)(ted)$_{0.5}$, the modes in the low frequency region are significantly affected by the hydration process: the doublet mode around 174 to 184 cm$^{-1}$ that involves the Cu-Cu vibration vanishes, and is replaced by a single broad peak at 187 cm$^{-1}$. The hydration process leads to the formation of a mode at ~ 268 cm$^{-1}$, which corresponds to the adsorption band of H$_2$bdc (See Fig. 6). In the high frequency region (950-1100 cm$^{-1}$ range), the C-C stretching mode $\nu_4$ is left unchanged at 1003 cm$^{-1}$. For Zn(bdc)(ted)$_{0.5}$, the doublet mode involving Zn-Zn species remains unaffected; the $\nu_4$ mode at 1018 cm$^{-1}$ disappears while a new band appears at 983 cm$^{-1}$. For Ni(bdc)(ted)$_{0.5}$, the Raman spectra do not significantly change upon hydration. For Co(bdc)(ted)$_{0.5}$, both the $\nu_4$ and the doublet modes disappear after hydration.

### 3.3 X-ray diffraction pattern

The crystal structure for guest-free M(bdc)(ted)$_{0.5}$ after activation includes 2D square networks, in which the paddle wheel SBUs[M$_2$(COO)$_4$] is linked by bdc ligands within the layer of the square networks (xy plane). The z axial sites of the metal ions are bonded by ted molecules to generate the 3D frameworks. For Cu(bdc)(ted)$_{0.5}$, all the phases except [001] shift to a higher 2θ value after hydration (See Fig. 8). For Zn(bdc)(ted)$_{0.5}$, the framework transforms into another phase after exposure to 9.5 Torr D$_2$O vapor for 1 h, in a manner similar to MOF-2 as evidenced by several characteristic peaks in Fig. S7. For Ni(bdc)(ted)$_{0.5}$, the powder X-ray pattern is not affected at all by hydration. In contrast, for Co(bdc)(ted)$_{0.5}$, the crystal structure is completely destroyed after exposure to 9.5 Torr D$_2$O vapor.

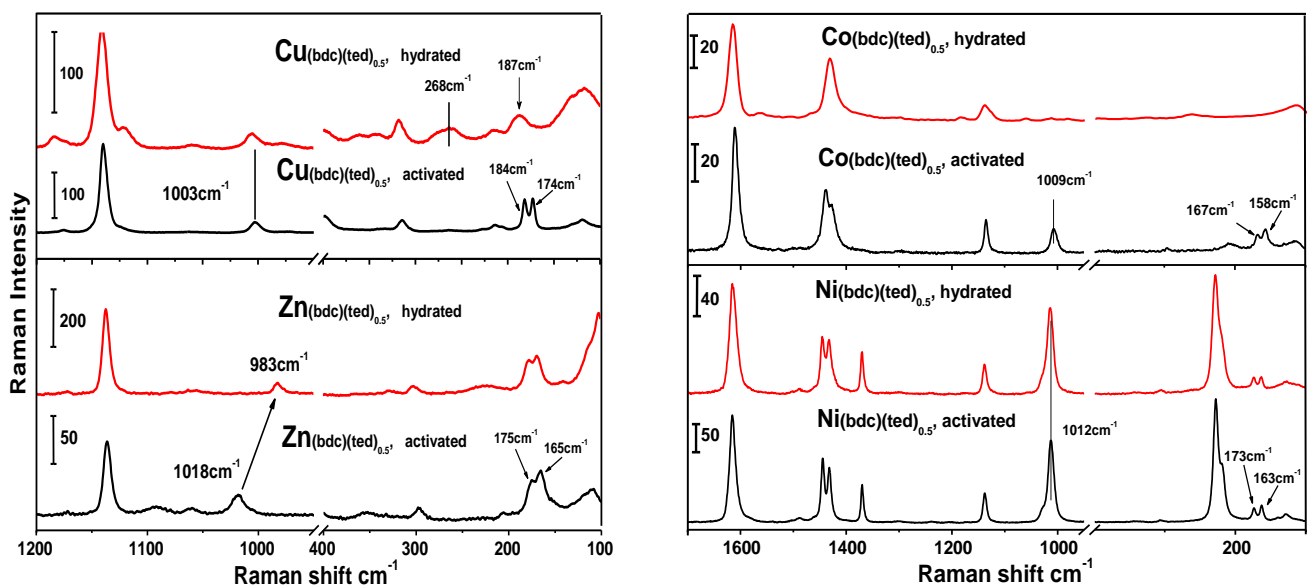

**Figure 7**. Raman spectra of activated (pristine) MOF samples and hydrated MOF materials after exposing to 9.5 Torr D$_2$O vapor.

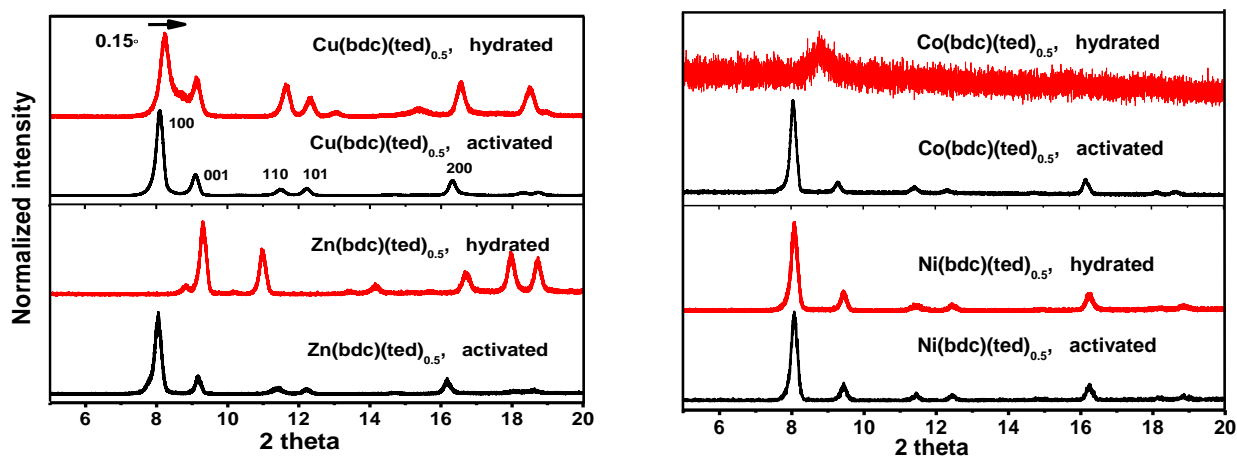

**Figure 8**. Powder X-ray pattern of hydrated MOF materials exposing to 9.5 Torr D$_2$O vapor and activated (pristine) MOF samples.

## 4 Interpretation

### 4.1 Cu(bdc)(ted)$_{0.5}$

For M(bdc)(ted)$_{0.5}$, the bdc and ted linkers are hydrophobic organic molecules surrounding a paddle wheel metal-oxide group. Hence, the adsorption sites for polar D$_2$O molecules are expected to be near the metal carboxylate.[73] This can explain that the $\nu_{asym}$(COO) and $\nu_{sym}$(COO) bands shift to lower wavenumbers when $D_2O$ is adsorbed into the frameworks. Powder X-ray diffraction measurements (See Fig. S10) confirm that Cu(bdc)(ted)$_{0.5}$ can be stable under low humidity (e.g. 6 Torr $D_2O$) without any major changes in the framework structure. The Raman spectra shown in Fig. S11 remain unchanged upon hydration by 6 Torr $D_2O$ exposure for 3hrs, which suggests that the MOF structure is stable under low humidity. The adsorbed $D_2O$ molecules interact with COO$^-$ but do not react with the framework. The sensitivity of the carboxylate mode to hydration and dehydration was mentioned in a previous study of UiO-66 metal organic frameworks.[74] The structure is maintained stable after 3-cycle absorption and desorption of water at the pressure up to 8 Torr (See Fig. S12). For each cycle, the absorption capacity for other gas (e. g. $CO_2$) is decreased upon pre-adsorbing water molecules. The absorption is recovered after regeneration of frameworks by removing the water molecules upon heating in vacuum. (See Fig. S12) Because of the favorable adsorption sites, we expect that M-O-C group is vulnerable to attack by $D_2O$ molecules. Our spectroscopic results indicate that, after exposing the activated sample to 9.5 Torr $D_2O$ vapor for 50 min, Cu(bdc)(ted)$_{0.5}$ is hydrolyzed by $D_2O$ molecules (Fig. 4) The direct experimental evidence is the appearance of the $\nu$(O-D) band at 2335 cm$^{-1}$, which can be ascribed to C-O-D due to deuteration of COO$^-$ by the $D_2O$ molecules. This vibrational assignment is consistent with that of the OD stretching band of deuterobenzoic acid monomers calculated at $\nu$ = 2333 cm$^{-1}$ using AM1 semiempirical methods and $w_{obs}$ = 2331 cm$^{-1}$ measured in Ar matrices.[75] Another weaker 2663 cm$^{-1}$ band may be associated with the $\nu$(O-D) bonded to Cu(II) ions.[76] The formation of a COOD group is also confirmed by three readily identified adsorption bands of COOD group at 1367 cm$^{-1}$, 1057 cm$^{-1}$ and 667 cm$^{-1}$. 1367 cm$^{-1}$ band must be due to C-O stretching mode while 1057 cm$^{-1}$ band and 667 cm$^{-1}$ band of a considerable breadth should be assigned to in-plane and out-of-plane OD deformation modes of dimeric carboxylic acids which are usually in the range of 675 ±25 cm$^{-1}$.[77] The assignment of the bands at 1367 cm$^{-1}$ and 1057 cm$^{-1}$ needs to take into account the coupling between the C-O stretching and OD deformation modes.[77] Another mode at 744 cm$^{-1}$ (shifting to 732 cm$^{-1}$) occurs at the frequency characteristic of free bdc acid ligands, and is therefore attributed to the benzene trigonal ring deformation mode represented by Wilson notation 12 in the Table 1.[54] Other minor bands at 1434, 1297 and 1113 cm$^{-1}$ correspond to the COOH group formed by protonation of COO$^-$ with $H_2O$ impurities from chamber and $D_2O$ source.[77] In all these measurements, the kinetic limitations are taken into account. For instance, we have monitored the $\nu$(O-D) band as a function of time, as shown in Fig. S14. Upon introduction of 9.5 Torr $D_2O$, the intensity of the $\nu$(O-D) bands reaches saturation in Cu(bdc)(ted)$_{0.5}$ after 1 h, and for Ni(bdc)(ted)$_{0.5}$, Co(bdc)(ted)$_{0.5}$, 2.5 h and 3 h, respectively. With this knowledge, sufficient time was given for each system to make sure that all reactions/adsorption were completed.

The Raman spectra obtained for Cu(bdc)(ted)$_{0.5}$ after hydration provide information any change associated with the local paddle wheel copper oxide cluster. As mentioned above, the doublet mode in the range of 174 to 184 cm$^{-1}$ involves Cu-Cu species and the 1003 cm$^{-1}$ band can be correlated to the C-C stretching mode $\nu_4$. The variations of the doublet mode shown in Fig. 7 are consistent with a reduction of the Cu-Cu interaction and an increase of the Cu-O bond strength due to formation of a Cu-OD group.[47] In the high frequency region (950-1100 cm$^{-1}$ range), the C-C stretching mode $\nu_4$ at 1003 cm$^{-1}$ is left unchanged, indicating that the ted molecules maintain the coordination to the Cu$^{2+}$ in the paddlewheel building units during the hydration process. The X-ray diffraction pattern clearly points to a structural change after hydration. All peaks shift to a higher 2θ value except for the [001], indicating that the distance between the 2D Cu$_2$(COO)$_4$ layers remains unchanged, which is consistent with the Raman observation that ted still coordinates to the metal sites after hydration. The two-theta position of the phases (i.e. interplanar distances) is controlled by the length of the bdc linkers and changes when these linkers are deuterated upon $D_2O$ exposure as shown in Fig. 8. From the XRD data, it is possible to conclude that the activated form is only partially hydrolyzed and is not collapsed after exposure to 9.5 Torr $D_2O$ vapor. The MOF structure still maintains its 3D network under 9.5 Torr vapor pressure. However, in liquid water, X-ray diffraction measurements have previously shown that the crystal structure is completely destroyed.[78] Even though it is only hydrolyzed, the framework structure cannot be regenerated by evacuation of water at higher temperature up to 150°C. (See Fig. S13)

### 4.2 Zn(bdc)(ted)$_{0.5}$

Chen has reported that air exposed Zn(bdc)(ted)$_{0.5}$ can be transformed into the two-dimensional structure MOF-2 [Zn(bdc)(H$_2$O)] and that MOF-2 can also be transformed back to Zn(bdc)(ted)$_{0.5}$ after adding the ted linkers to MOF-2 in DMF at 110°C for 2 days[79] Both Zn(bdc)(ted)$_{0.5}$ and MOF-2 contain the similar square-grid Zn$_2$(bdc)$_2$ composed of di-nuclear paddlewheel Zn-Zn units bridged by bdc di-anions. However, in MOF-2, the square-grid layers are held together by hydrogen bonding between water and two adjacent paddlewheels, with the oxygen of H$_2$O bound to the Zn atom and the H of the same H$_2$O molecule bound to the carboxylate oxygen of the adjacent paddlewheel structure.[80, 81] The XRD data of Zn(bdc)(ted)$_{0.5}$ shown in Fig. 8 clearly shows that the frameworks transform into MOF-2 after hydration. This transformation first involves the detachment of ted molecules from the frameworks, then the bonding of $D_2O$ molecules to Zn$^{2+}$ apical sites of the paddle wheel building units through their oxygen atoms. An in-situ IR study confirms this process as shown below by examination of water related modes.

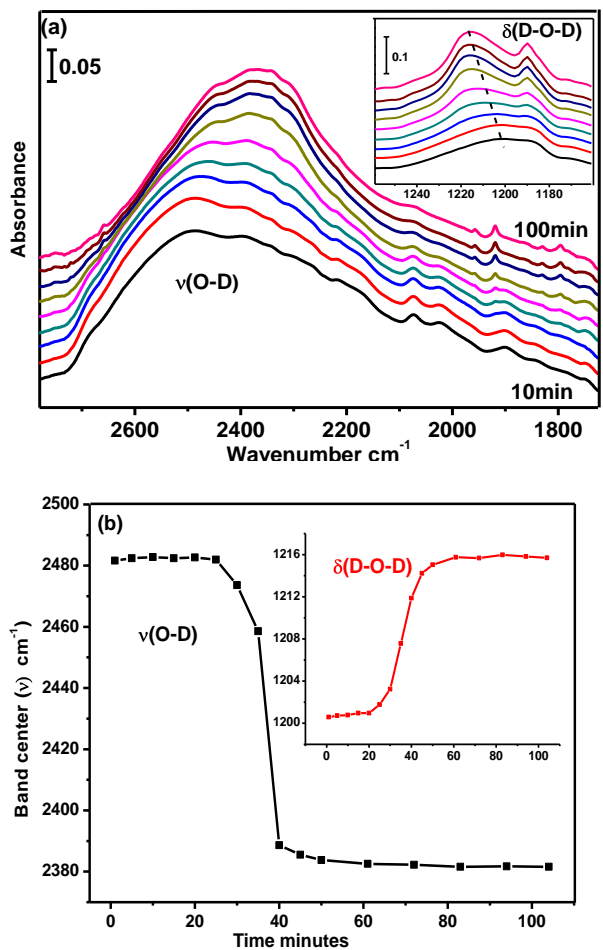

**Figure 9.** (a) Time dependent IR spectra of $D_2O$ adsorption into $Zn(bdc)(ted)_{0.5}$ over 100 min, all referenced to activated $Zn(bdc)(ted)_{0.5}$ in vacuum. Inset shows bending mode $\delta(D\text{-}O\text{-}D)$ change as a function of time. From bottom to top: 10, 20, 30, 35, 40, 50, 60, 80, 100 min. (b) $\nu(O\text{-}D)$ band center change over time for 100 min and inset shows the bending mode change.

Instead of seeing the formation of a sharp $\nu(O\text{-}D)$ band between 2700 and 2600 cm$^{-1}$ by protonation reaction after exposure to $D_2O$ vapor for the Cu, Co and Ni MOFs (See Fig. 4), there is only a broad and red-shifted $\nu(O\text{-}D)$ band and a blue-shift shifted $\delta(D\text{-}O\text{-}D)$ band in $Zn(bdc)(ted)_{0.5}$, which both slowly increase with time after introduction of 9.5 Torr vapor $D_2O$ as shown in Fig. 10. Initially, the $\nu(O\text{-}D)$ band is centered at 2485 cm$^{-1}$, characteristic of $D_2O$ adsorption band for the other three $M(bdc)(ted)_{0.5}$ compounds at 9.5 Torr (See Fig. S15). The $\delta(D\text{-}O\text{-}D)$ band is centered at 1200 cm$^{-1}$ due to $D_2O$ condensation, as discussed later. After 1h, the $\nu(O\text{-}D)$ band center gradually shifts to lower wavenumber (~2382 cm$^{-1}$) and the $\delta(D\text{-}O\text{-}D)$ band shifts to a higher wavenumber (~1216 cm$^{-1}$). These shifts for $\nu(O\text{-}D)$ and $\delta(D\text{-}O\text{-}D)$ indicate that the deuterium bonding of the $D_2O$ molecules is stronger.[82-84] These shift are only observed in $Zn(bdc)(ted)_{0.5}$ at higher vapor pressure (~9.5 Torr) after $D_2O$ condensation adsorption. In other three isostructural MOFs, there are no observable shifts as a function of time. In MOF-2 structures, a bifurcated deuterium bond exists between the $D_2O$ molecules and O atoms of carboxylate groups and O atoms to the Zn metals in adjacent 2D layers.[81] This type of configuration can explain the $D_2O$ vibration mode shift and the high thermal stability of hydrated $Zn(bdc)(ted)_{0.5}$ in Fig. 5.

The Raman spectra of hydrated $Zn(bdc)(ted)_{0.5}$ shown in Fig. 7 can be used to determine the environment of the N lone pairs in the ted ligands, by utilizing the sensitivity of the $\nu_4$ C-C stretching mode the state of the N atom. The C-C stretching mode of ted molecules in the activated $Zn(bdc)(ted)_{0.5}$ is at 1018 cm$^{-1}$, and red shifts by 35 cm$^{-1}$ to 983 cm$^{-1}$ after hydration, which suggests that the nitrogen atom of the ted molecules is no longer connected to the metal atoms. This indicates the removal of ted molecules from the apical site of paddle wheel Zn clusters by $D_2O$ molecules. In the low frequency region (100 to 400 cm$^{-1}$), the doublet mode associated with the Zn-Zn bonding persists (See Fig. 7), indicating that the paddle wheel structure remains unbroken by $D_2O$ molecules.

### 4.3 Ni(bdc)(ted)$_{0.5}$

Powder X-ray diffraction results and Raman spectra show that the crystal structure after $D_2O$ exposure remains intact in Fig. 7 and Fig. 8. After exposing $Ni(bdc)(ted)_{0.5}$ to 9.5 Torr $D_2O$ vapor for 2.5 h, an OD stretching band appears at 2660 cm$^{-1}$ in Fig. 4. This band cannot be assigned to COOD for several reasons: (i) for the isostructural compound with $Cu(bdc)(ted)_{0.5}$, in which the bdc molecules is deuterated, the $\nu(OD)$ band is observed at 2635 cm$^{-1}$; (ii) the three characteristic adsorption bands for COOD in the region from 1500 to 500 cm$^{-1}$, including C-O stretching, in plane and out-of-plane OD deformation at 1367, 1057, and 667 cm$^{-1}$ are absent; and yet (iii) the benzene ring deformation mode $\sigma_{12}$ at 744 cm$^{-1}$ indicates that $D_2O$ is definitely adsorbed into the frameworks. In the case of $Cu(bdc)(ted)_{0.5}$ shown in Fig. 10(a), this mode shifts 12 cm$^{-1}$ to 756 cm$^{-1}$ upon exposure to 8.5 Torr or at the beginning of the 9.5 Torr exposure. When the bdc linkers are deuterated under higher $D_2O$ vapor pressure, it shifts back to 732 cm$^{-1}$ with time. In $Ni(bdc)(ted)_{0.5}$, this mode remains at high frequencies with 12 cm$^{-1}$ and 8 cm$^{-1}$ shifts upon $D_2O$ adsorption in all the pressure ranges (See Fig. 10 and Fig. S15). The red shift of the structure-sensitive band $\nu_{18a}$ at 1017 cm$^{-1}$ is also different from what is observed in $Cu(bdc)(ted)_{0.5}$. There is therefore no clear evidence for dissociation of the bond between the metal oxide clusters and bdc linkers in IR and Raman spectroscopy.

With regard to the COO vibrations and ring deformation modes described in the result section (Table 1 and Fig. 4), there are important differences in their response to water molecules that underlie their reaction characteristics, as summarized in Fig. 10. In this figure, two $D_2O$ pressures are chosen, 8 Torr and 9.5 Torr because the structures are still stable at 8 Torr, but react at 9.5 Torr. An examination of the $\nu_{sym}(COO)$ band for the four $M(bdc)(ted)_{0.5}$ compounds during the time dependence experiment at 9.5 Torr in Fig. 10 reveals that there are significant differences between $Ni(bdc)(ted)_{0.5}$ and $Cu(bdc)(ted)_{0.5}$ and $Zn(bdc)(ted)_{0.5}$. For $Cu(bdc)(ted)_{0.5}$. At 8 Torr, the features in the adsorption spectra of Fig. 10 (a) are dominated by perturbations bands caused by incorporation of $D_2O$ molecules, similarly to what is observed at 6 Torr in Fig. 3. When the pressure is increased to 9 Torr, some changes occur with time in the absorption features at 1367 cm$^{-1}$, 667 cm$^{-1}$ as well as 1434, 1297, 1113 cm$^{-1}$. The formation of these features is the result of carboxylate (COO$^-$) group reaction with D$_2$O molecules and trace amount of H$_2$O molecules. As discussed above, the bands at 1367 cm$^{-1}$ and 667 cm$^{-1}$ are due to $\nu$(C-O)+$\delta$(O-D) and out of plane $\delta$(CO-D) respectively. The 1434, 1297, and 1113 cm$^{-1}$ bands are associated with the COOH group. After evacuation at 373 K, the $\nu$(C=O) band in Cu(bdc)(ted)$_{0.5}$ becomes observable because it sharpens and blue shifts back to 1685 cm$^{-1}$ as the deuterium bonding to oxygen atoms of C=O group is released.

For Zn(bdc)(ted)$_{0.5}$, the perturbed $\nu_{sym}$(COO) band at 1361 cm$^{-1}$ observed at 8.0 Torr further red shifts to 1353 cm$^{-1}$ after the displacement reaction of D$_2$O with ted linkers slowly taking place at 9.5 Torr. This shift is indicative of a stronger host-guest interaction between the carboxylate moieties and the D$_2$O molecules. This result is consistent with the red shift of $\nu$(O-D) and blue shift $\delta$(D-O-D) of the adsorbed D$_2$O shown in Fig. 10(b). This red shift cannot be recovered by evacuating at 373 K because the D$_2$O molecules remain bifurcated-deuterium bonded with the COO group shown in Fig. 5 and in Fig. 10(b) for the bending mode. In the low frequency region (600 to 700 cm$^{-1}$), there is no evidence for the formation of out of plane OD deformation mode, confirming that the Zn-O-C group was not broken during the D$_2$O exposure.

For Ni(bdc)(ted)$_{0.5}$, the perturbed $\nu_{sym}$(COO) band at 1364 cm$^{-1}$ at 8 Torr shifts back to a higher value of 1378 cm$^{-1}$ closer to its original position and the $\nu_{asym}$(COO) shifts from 1567 to 1574 cm$^{-1}$ when the vapor pressure reaches 9.5 Torr, which indicates that the deuterium bonding between the COO group and the guest D$_2$O molecules is greatly weakened. The perturbed $\nu_{sym}$(COO) band disappears when adsorbed D$_2$O molecules are removed by heating at 373 K under vacuum as shown in Fig. 10(c). However, in the case of Zn(bdc)(ted)$_{0.5}$ and Cu(bdc)(ted)$_{0.5}$, the bands at 1352 cm$^{-1}$ and 1367 cm$^{-1}$, arising from the deuteration reaction, persist even after evacuation at elevated temperature. The question remains how to explain the formation of the isolated $\nu$(O-D) band at 2660 cm$^{-1}$ in Fig. 4. The above IR spectroscopic results and analysis clearly show that the bdc linkers are not deuterated by D$_2$O molecules. Similarly, the Raman and XRD results (See Fig. 7 and Fig. 8) show that the structure of Ni(bdc)(ted)$_{0.5}$ remains intact after D$_2$O exposure. All the characteristic bands of Ni(bdc)(ted)$_{0.5}$ are preserved after hydration without addition of new bands in the Raman spectra. Therefore, it is more reasonable to attribute this isolated ν(O-D) band to D$_2$O molecules instead of OD group. The D$_2$O molecules may coordinate to the Ni(II) ions not by breaking Ni-O and Ni-N bonds within the MOF structure. In support of this hypothesis, the broad band at 625 cm$^{-1}$ in Fig. 4 and Fig. 10(c) is assigned to the out-of-plane (D-O-D) deformation mode. The scissor mode of such an isolated D$_2$O would occur between in the 1050 to 1270 cm$^{-1}$ region examined in Fig. 4 and Fig. 10(c): a broad peak around 1190 to 1203 cm$^{-1}$ is assigned to scissor mode of adsorbed deuterated water. In addition, there is a sharp band at 1105 cm$^{-1}$ that is not found at low pressure D$_2$O vapor exposure but appears together with ν(O-D) at 2660 cm$^{-1}$ at higher vapor pressure in Fig. 10(c). Although such a red shift is not usually observed for H-bonded water, previous studies[85-87] have shown that the H-O-H scissor frequency can be considerably red shifted in some crystalline hydrates compared to the position of δ(H-O-H) of free water molecules (1594.6 cm$^{-1}$). For example, tetrahedral coordination of the water molecules and formation of chains M-O$_w$-M in the kieserite family MSO$_4$·H$_2$O causes δ(H-O-H) to shift to a lower wavenumber by 100 cm$^{-1}$.[85] On the basis of these studies, we assign the 1105 cm$^{-1}$ band to be isolated adsorbed D$_2$O in Ni(bdc)(ted)$_{0.5}$. This D$_2$O is likely coordinated to the two Ni atoms in the dinuclear paddlewheel units. However, the degree of deuterium bonding with the O atoms of the COO group varies, leading to an apparent shift of the ν$_{sym}$(COO) band at 1364 cm$^{-1}$ toward 1380 cm$^{-1}$.

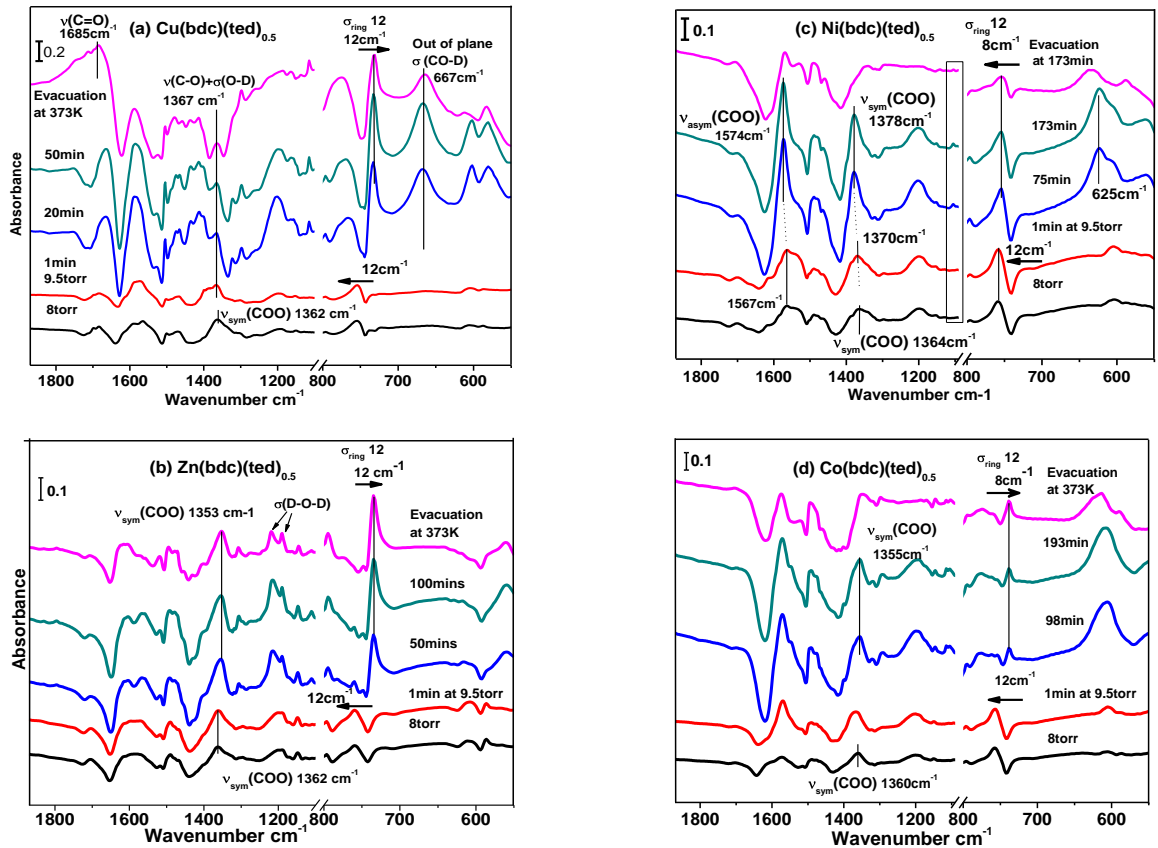

Figure 10. IR adsorption spectra of hydrated MOF during D$_2$O exposure, reference to the activated MOF in a vacuum. (a) Cu(bdc)(ted)$_{0.5}$: black, red, blue and dark green refer to after introduction of 8 Torr D$_2$O vapor for 40 min, after introduction of 9.5 Torr D$_2$O vapor for 1, 20, and 50 min; pink, evacuation at 373 K for 10 h. (b) Zn(bdc)(ted)$_{0.5}$: black, red, blue and dark green refer to after introduction of 8 Torr D$_2$O vapor for 40 min, after introduction of 9.5 Torr D$_2$O vapor for 1, 50, 100 min; pink, evacuation at 373 K for 10 h. (c) Ni(bdc)(ted)$_{0.5}$: black, red, blue, and dark green refer to after introduction of 8 Torr D$_2$O vapor for 40 min, after introduction of 9.5 Torr D$_2$O vapor for 1, 75, and 173 min; pink, evacuation at 373 K for 10 h. (c) Co(bdc)(ted)$_{0.5}$: black, red, blue and dark green refer to after introduction of 8 Torr D$_2$O vapor for 40 min, after introduction of 9.5 Torr D$_2$O vapor for 1, 98 and 193 min; pink, evacuation at 373 K for 10 h.

### 4.4 Co(bdc)(ted)$_{0.5}$

From the powder X-ray diffraction results (See Fig. 8), it is clear that the crystal structure of Co(bdc)(ted)$_{0.5}$ is completely destroyed after exposure to 9.5 Torr D$_2$O vapor. The framework structure cannot be recovered after annealing in vacuum up to 150 °C to remove adsorbed water. (See Fig. S13) To understand the dissociation of the MOFs structure, we examine the evolution of the spectroscopic data. As for Ni(bdc)(ted)$_{0.5}$, there are no bands associated with the COOD group in the region from 1500 to 500 cm$^{-1}$ in the IR adsorption spectra of hydrated Co(bdc)(ted)$_{0.5}$ in Fig. 4. Moreover, there is a red shift of ν$_{18a}$ from 1017 cm$^{-1}$ to 1013 cm$^{-1}$. These spectroscopic results indicate that there is no hydrolysis reaction with Co-O-C during D$_2$O vapor exposure. The bdc linkers are still coordinated with Co(II) ions by COO$^-$ group. However, the loss of the ν$_4$ mode and doublet mode at 158 and 167 cm$^{-1}$ in the Raman spectra of Fig. 7 indicates that the coordinated ted linkers disappear and the Co-Co structure is affected by hydration. By comparing Fig.10 (b) and (d), we find that the Co(bdc)(ted)$_{0.5}$ and Zn(bdc)(ted)$_{0.5}$ exhibit similar shifts of ν$_{sym}$(COO) and ring deformation bands upon D$_2$O inclusion. For Co(bdc)(ted)$_{0.5}$, the ν$_{sym}$(COO) mode at 1360 cm$^{-1}$ red shifts a little to a lower wavenumber of 1355 cm$^{-1}$. The ring deformation band σ$_{12}$ blue shifts by 12 cm$^{-1}$ at 8 Torr but red shifts by 8 cm$^{-1}$ at 9.5 Torr under hydration. These changes in the COO modes are similar in the spectra of hydrated Co(bdc)(ted)$_{0.5}$ and Zn(bdc)(ted)$_{0.5}$. However, the D$_2$O bands including ν(O-D) and δ(D-O-D) behave differently for these two MOFs. The shift of ν(O-D) and δ(D-O-D) in Zn(bdc)(ted)$_{0.5}$ mentioned before is result of displacement of ted molecules by D$_2$O. For Co(bdc)(ted)$_{0.5}$, a isolated ν(O-D) band is identified at 2356 cm$^{-1}$ after hydration shown in Fig. 4 and remains after evacuation at 373 K for 10 h as shown in Fig. 5. We therefore conclude that D$_2$O molecules attack the Co-N bond and replace the ted linker to bond the apical Co(II) site. However, unlike the D$_2$O molecules in hydrated Zn(bdc)(ted)$_{0.5}$ that have a bifurcated-deuterium bonding with Zn$_2$(bdc)$_2$ layers, the D$_2$O molecules in hydrated Co(bdc)(ted)$_{0.5}$ are coordinated to Co(II) ions via one oxygen atom only. As a result, the paddle wheel Co$_2$(COO)$_4$ is distorted and the Co-Co interaction is removed. This is why the doublet involving Co-Co in Raman spectra of Fig. 8 disappears in hydrated Co(bdc)(ted)$_{0.5}$ and the crystal structure collapses after hydration. Consequently, we ascribe the broad band at 612 cm$^{-1}$ in Fig. 4 and Fig. 10(d) to the out-of-plane (D-O-D) deformation mode.

### 5 Discussion

#### 5.1. Condensation

To study the water adsorption behavior, we performed the pressure dependence measurement of D$_2$O vapor adsorption into M(bdc)(ted)$_{0.5}$. The spectra were recorded as a function of vapor pressure from ~1 Torr to ~9.5 Torr with 40 min to equilibrate at each pressure. Fig. 11 shows 1) the spectra upon D$_2$O adsorption into Cu(bdc)(ted)$_{0.5}$ in the range of the ν(O-D) and σ(D-O-D) modes and 2) the integrated areas of these two modes as a function of pressure up to 9.5 Torr. In the low pressure range (up to 8 Torr), the integrated areas of both ν(O-D) and σ(D-O-D) modes increase almost linearly with pressure. A steep increase occurs around 8.5 Torr, which is assumed to be the result of D$_2$O molecules condensation in the pore. The inset in Fig. 11(a) provides spectroscopic evidence for condensation by the shift of σ(D-O-D) with increasing pressure from 1189 cm$^{-1}$ to 1200 cm$^{-1}$, which is the typical frequency of σ(D-O-D) in the liquid D$_2$O.[84] In Fig. 12, we can also see that D$_2$O adsorption increases as the temperature is lowered, as is expected for physisorption in micropores.

Fig. S15 presents the infrared absorption spectra for M(bdc)(ted)$_{0.5}$ [M=Zn, Ni, Co] as a function of D$_2$O pressure. All of them show the same linear increase in low pressure region and a steep increase at 8.5 Torr as observed in Cu(bdc)(ted)$_{0.5}$, indicating that the physical D$_2$O adsorption in the M(bdc)(ted)$_{0.5}$ series is independent of the central metal ions.

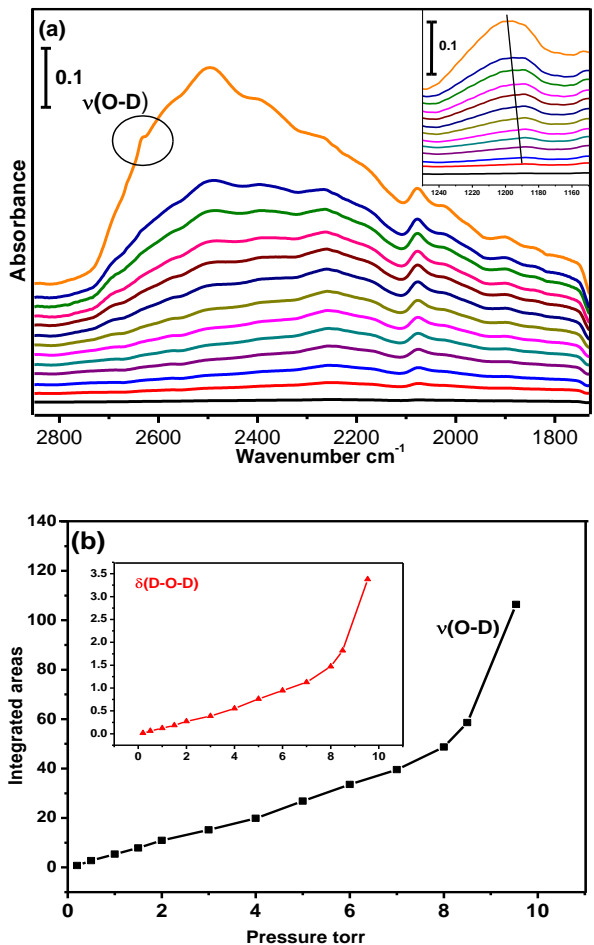

**Figure 11.** (a), IR spectra of $D_2O$ adsorption into $Cu(bdc)(ted)_{0.5}$ as a function of pressure from bottom to top: 200 mTorr, 500 mTorr, 800 mTorr, 1, 2, 3, 4, 5, 6, 7, 8, 8.5, 9.5 Torr. Inset shows the bending mode region from 1150 to 1250 cm$^{-1}$. All referenced to activated MOF in vacuum. (b), integrated areas of stretching band in the region of 2736-1753 cm$^{-1}$ as a function of vapor pressure. Inset shows integrated areas of bending mode as a function of pressure.

In all four compounds, we see that the reaction of $D_2O$ with the paddle wheel structure is initiated upon condensation of $D_2O$ inside the pores, initiates reaction as shown in Fig. 11 and Fig. S15. For example, in $Cu(bdc)(ted)_{0.5}$, the Cu-O-C group is hydrolyzed into Cu-OD and C-OD as $D_2O$ is dissociated,. This reaction does not take place at low $D_2O$ vapor pressure. In this low pressure regime, the $D_2O$ molecules do not induce decomposition of frameworks, although the MOF structure is perturbed, as evidenced by strong variations in the MOF phonon bands. The reaction of $D_2O$ with the metal oxide bonds in the MOF is reminiscent of water adsorption on metal oxide surfaces, such as MgO, ZnO, and NiO, for which there are theoretical predictions of strong dependence of the dissociation barrier on water coverage.[88-90] A First-principles molecular-dynamics simulation has predicted that an isolated water molecule on MgO[100] surface is not sufficient to induce dissociation, only bonding molecularly through one hydrogen of the water molecule to a surface oxygen. Above 1/2 ML coverage, the dissociation of waters occurs and a mixture of molecular and dissociated water molecules is observed at 2/3 ML and 1 monolayer coverage. This can be explained by the fact that a hydrogen bond is formed between water molecules and that this hydrogen bond weakens the OH bond in neighboring molecules, lowering the barrier for water dissociation. The same situation takes place for adsorbed $D_2O$ molecules in $Cu(bdc)(ted)_{0.5}$ compounds: the condensation of $D_2O$ molecules into the framework weakens the OD bonds and induces hydrolysis with the Cu-O-C group. (See Fig.11) The temperature dependence experiments confirm this picture in which condensation is critical for reaction. At higher temperature of 50 °C and 40 °C, the framework is stable to 13 Torr $D_2O$ vapor because condensation does not occur. When the temperature is decreased to 35 °C, the $D_2O$ molecules begin to react with frameworks as evidenced by the appearance of the ν(O-D) band at 2635 cm$^{-1}$ and 2663 cm$^{-1}$ that confirms a deuteration reaction of the bdc linkers with the $D_2O$ molecules. At the same time, the center of $D_2O$ bending mode shifts to 1200 cm$^{-1}$ (see inset b of Fig. 12), confirming that $D_2O$ molecules are condensing into the frameworks at 35 °C.

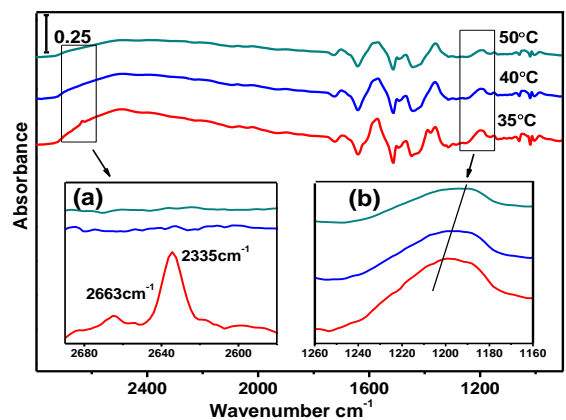

**Figure 12.** IR spectra of $D_2O$ adsorption into $Cu(bdc)(ted)_{0.5}$ with temperature decreasing from 50 °C, 40 °C to 35 °C at constant pressure of 13 Torr $D_2O$ vapor, referenced to activated MOF in vacuum. Inset a and b shows stretching mode region and bending mode region.

**5.2 Effects of Metal Ions on Stability and Reaction Pathway:**

Huang *et al.* first reported that exposure of MOF-5 to water resulted in possible hydrolysis of the materials and the formation of terephthalic acid ($H_2bdc$).[20] Later, Greathouse and Allendorf used empirical force fields and molecular dynamics to predict that this reaction is initiated by a direct attack of MOF-5 by water molecules.[21] The weak bonds between Zn and O atoms in MOF-5 can be broken when interacting with water molecules. Low *et al.* used a quantum mechanical to model the hydrolysis reaction of water with metal oxide clusters, showing that it involved the breaking and reforming of bonds in different types of metal organic frameworks depending on the specific secondary building units (SBUs).[18] The proton of water molecules can attack the oxygen of metal oxide cluster when the carboxylate opens up a coordination site on the metal for the water molecules. Later, Han used a reactive force field (ReaxFF) approach to perform a detailed study of hydrolysis of MOF-5 and found a direct water interaction with $ZnO_4$ tetrahedron in framework induces cleavage of Zn-O bonds and dissociation of $H_2O$ molecules of into OH and H. The resulting OH forms a chemical bond with Zn, and the proton is attached to the oxygen of the bdc moiety.[23]

From our experimental results of the four isostructural compounds M(bdc)(ted)$_{0.5}$ [M=Cu, Zn, Ni, and Co] with paddlewheel dinuclear metal clusters, we find that both the initial decomposition pathway and the stability of the MOF frameworks with respect to reaction with $D_2O$ molecules are dependent on the central metal ions. For Cu(bdc)(ted)$_{0.5}$, the hydrolysis reaction of $D_2O$ molecules with metal oxide clusters Cu-O-C can be confirmed by our *in-situ* IR spectroscopy measurements under high humidity conditions (48%). However, the hydrolysis decomposition mechanism does not hold true in all isostructural compounds of M(bdc)(ted)$_{0.5}$ under the same conditions. For Zn(bdc)(ted)$_{0.5}$, the Zn-N bonds are broken by reaction with $D_2O$ molecules so that ted molecules are released from the apical sites of the paddlewheel dinuclear $Zn^{2+}$ clusters and the $D_2O$ molecules bond to Zn SBUs. In contrast, the Cu-N group in Cu(bdc)(ted)$_{0.5}$ structure is less susceptible to displacement by incoming water molecules because the overall formation (stability) constant of its amine complex in aqueous solution. Both Cu(II) and Zn(II) form amine complexes, Cu(NH$_3$)$_4^{2+}$ and Zn(NH$_3$)$_4^{2+}$, in an ammonia medium, although with very different stability constants: $1.1 \times 10^{13}$ and $2.8 \times 10^9$, respectively.[91]

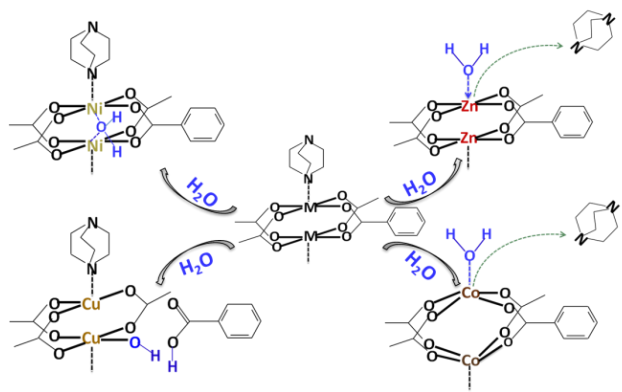

**Figure 13.** Schematic illustration of decomposition pathway of M(bdc)(ted)$_{0.5}$ [M=Cu, Zn, Ni, Co] reaction with $D_2O$ molecules.

Ni(bdc)(ted)$_{0.5}$ is more stable in $D_2O$ vapor than Cu(bdc)(ted)$_{0.5}$ and Zn(bdc)(ted)$_{0.5}$, at least under the same condition of 9.5 Torr vapor exposure at RT as shown from XRD measurements in Fig. 8. The Raman spectra after $D_2O$ exposure also confirm that the metal-oxide paddle-wheel did not break down. The relatively higher stability of Ni(bdc)(ted)$_{0.5}$ against hydrolysis reaction is consistent with a model of water interaction with the surface of rocksalt NiO (100). Indeed, there is a consensus both from experimental investigations and theoretical calculations that water adsorbs molecularly on NiO(100) surfaces and that a dissociation reaction can only take place at defect sites.[92-95] Simpson's theoretical calculation by using the semi-empirical SCFMO method MSINDO shows that water dissociation is unlikely on the planar surface of NiO(100) and the associated activation energy for dissociation is high due to the rigidity of the NiO(100) lattice which prevents water molecule from adopting a stable transition state,[96] suggesting by analogy that Ni(bdc)(ted)$_{0.5}$ is more stable in water vapor. These conclusions are also supported by considering the metal-oxygen strength from the dissociation energy for diatomic molecules: Zn-O(<250.4 kj/mol), Cu-O(~287.4 kj/mol), Ni-O(~366 kj/mol),Co-O(~397.4 kj/mol).[97] It is therefore expected that Ni(bdc)(ted)$_{0.5}$ is more stable than Cu(bdc)(ted)$_{0.5}$ to hydrolysis reaction with water molecules. The stability of the Ni-N group of ted linkers to a displacement reaction by incoming water molecules can be also justified by the overall formation (stability) constant of the hexaaminemetal complex in aqueous solution. The constant of Ni(NH$_3$)$_6^{2+}$ is $2.0 \times 10^8$, much higher than that of Co(NH$_3$)$_6^{2+}$ $2.0 \times 10^4$.[91] It is experimentally observed that cobalt(II) nitrate hydroxide precipitates in an ammonia medium rather than forming amine complexes,[98] which is consistent with our observation that Co-N in Co(bdc)(ted)$_{0.5}$ is easily displaced by $D_2O$ molecules. The displacement reaction induces the Co(bdc)(ted)$_{0.5}$ structure collapse under $D_2O$ vapor. However, the high Co-O bond strength (397.4 kj/mol) prevents the bdc moieties from being deuterated by $D_2O$ molecules. Fig. 13 presents a schematic illustration of the initial decomposition pathway of M(bdc)(ted)$_{0.5}$ [M=Cu, Zn, Ni, Co] reaction with $D_2O$ molecules at 9.5 Torr. For Ni(bdc)(ted)$_{0.5}$, if we slightly increase the pressure to 9.7 Torr, condensation occurs the framework structure is broken (See Fig. S16). This result is consistent with Liang's study that shows that the crystal structure of Ni(bdc)(ted)$_{0.5}$ partially collapses at high humidity levels (e g., ~60%) at RT.[64] However, the order of stability of M(bdc)(ted)$_{0.5}$ series does not completely follow the empirical Irving-Williams sequence for relative stabilities of metal complexes (i.e., "Mn$^{2+}$<Fe$^{2+}$<Co$^{2+}$<Ni$^{2+}$<Cu$^{2+}$>Zn$^{2+}$").[99]

### 5.3 Theoretical investigation

Aiming to understand how water can influence the M(bdc)(ted)$_{0.5}$ structure, we performed vdW-DF simulations. We followed the scheme shown in Fig. 14, where water molecules replaced the ted group.

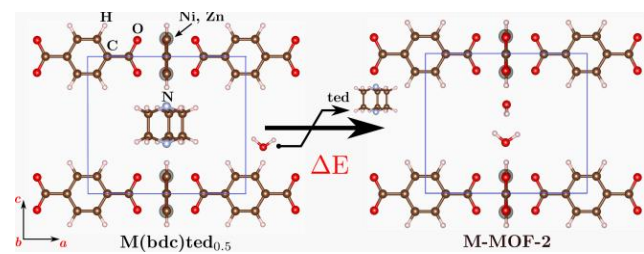

**Figure 14.** Scheme adopted for water insertion in 2·M(bdc)(ted)$_{0.5}$ [M=Zn, Ni], where the ted group has been substituted by some water molecules.

The relevant physical quantity governing the hydration of M(bdc)(ted)$_{0.5}$ is $\Delta E$ (See Fig. 14), which represents the energy necessary to extract and replace the ted unit with water molecules:

$$\Delta E = [E(2 \cdot M(bcd)ted_{0.5}) + nE(W) - E(M\text{-MOF-2 } nH_2O) - 1/2E(ted)] \quad 1$$

where **n** is the number of water molecules introduced. M-MOF-2 $nH_2O$ is the $M(bdc)(ted)_{0.5}$ where ted was replaced by **n** water molecules. Initially, only two water molecules (per cell) were introduced, coordinating the bare metal sites [M=Zn and Ni]. The number of water molecules (per structure) was then progressively increased up to 10 molecules (per cell, 5 per $M(bdc)(ted)_{0.5}$ unit), simulating the effect of increasing water pressure. Fig. 15 shows the oxygen local environment of the metal atoms [M=Zn and Ni] available in the $2 \cdot M(bdc)(ted)_{0.5}$ for H-bridging with incoming water molecules.

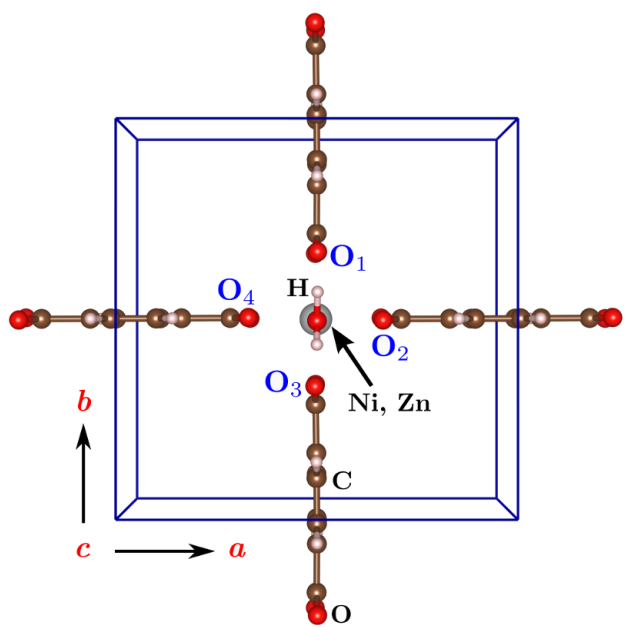

**Figure 15.** View along the [0 0 1] plane of the $M(bdc)(ted)_{0.5}$ (M-MOF-2) [M=Zn and Ni] once the ted unit has been replaced by two water molecules (here only one water molecule is shown, see Figure 14). Labels $O_1$, $O_2$, $O_3$, and $O_4$ show the local environment of the metal sites Ni, Zn, on which water molecules were progressively adsorbed, simulating the effect of increasing water pressure

**Table 2.** Computed $\Delta E$ and $\Delta E_{MOF-2}$ (kJ mol$^{-1}$ cell$^{-1}$) as a function of the number of water molecules per cell. $|\Delta|$ is the difference between $\Delta E$ (or $\Delta E_{MOF-2}$) for Zn and Ni.

| No. $H_2O$/cell | $\Delta E$ | | | $\Delta E_{MOF-2}$ | | |
|---|---|---|---|---|---|---|
| | Zn | Ni | $|\Delta|$ | Zn | Ni | $|\Delta|$ |
| 2 | +43.1 | +85.5 | 42.4 | 0.0 | 0.0 | 0.0 |
| 4 | -5.3 | +4.2 | 9.5 | -53.6 | -77.1 | 23.5 |
| 6 | -21.4 | -17.1 | 4.3 | -53.7 | -68.4 | 14.7 |
| 8 | -31.3 | -24.0 | 7.3 | -56.1 | -52.4 | 3.7 |
| 10 | -35.0 | -45.2 | 10.2 | -54.5 | -55.3 | 0.8 |

At first glance, the $\Delta E$'s values calculated for Zn and Ni follow the trend found experimentally, i.e. water seems less reactive with $Ni(bdc)(ted)_{0.5}$. On the other hand, the hydration of $Zn(bdc)(ted)_{0.5}$ is a highly spontaneous process even when only a few water molecules (4 molecules, See Table 2) are introduced in the unit cell. Experimentally, the sorption capacity of $Zn(bdc)(ted)_{0.5}$ was estimated at approximately 2.5 molecules per unit cell, which is in excellent agreement with results of water vapor sorption isotherm measurement.[62] Here the driving force is the formation of strong H-bonds between water molecules, which stabilizes the coordination of Zn metal sites. Although strong H-bonding networks also take place in the case of $Ni(bdc)(ted)_{0.5}$, this does not decrease the $\Delta E$ to negative values, hence explaining our experimental findings. Alternatively we can imagine the hydration of the M-MOF structure (where this latter is formed when n =2 in Eq. 1):

$$\Delta E_{MOF-2} = [E(M\text{-MOF-2}) + nE(W) - E(M\text{-MOF-2 } nH_2O)] \quad 2$$

The hydration of both Zn- and Ni-MOF-2 is a spontaneous process, and this is slightly more favorable for Ni. For situations of high loading (n = 10) $\Delta E_{MOF-2}$ reaches a steady value of 55 kJ mol$^{-1}$ cell$^{-1}$ (see Table 2). However, $\Delta E_{MOF-2}$ measures only the strength of the hydrogen-bond network acting between water molecules.

## 6. Conclusion

In summary, we have experimentally investigated the interaction of $D_2O$ with prototypical metal organic frameworks $M(bdc)(ted)_{0.5}$ [M=Cu, Zn, Ni and Co]. We find that the structural stability of $M(bdc)(ted)_{0.5}$ compounds is determined by the water content inside the MOF. At lower loading, the structures remain intact. At higher loading, due to water condensation, the structures can decompose by reaction of water with the paddle wheel metal cluster and organic linkers, as observed by *in-situ* IR spectroscopy and *ex-situ* Raman scattering. Condensation was determined through pressure and temperature dependence studies. We also find that the initial decomposition pathways sensitively depend on the central metal ions: For $Cu(bdc)(ted)_{0.5}$, a hydrolysis reaction of water molecules with Cu-O-C group induces the paddle wheel structural decomposition. For $Zn(bdc)(ted)_{0.5}$ and $Co(bdc)(ted)_{0.5}$, the water molecules replace ted pillars and bond to the apical sites of paddle wheel $Zn_2(COO)_4$ and $Co_2(COO)_4$. Finally, we find that the overall stability of isostructural MOFs $M(bdc)(ted)_{0.5}$ follows the order of Cu-MOF<Ni-MOF>Zn-MOF> Co-MOF and this order can be correlated with bond dissociation energy of diatomic molecules metal-oxygen and overall formation(stability) constants of metal amine complexes. The findings of this work are supported by first-principles calculations, providing the information necessary for determining operating conditions of this class of MOFs with a paddle wheel secondary building unit in harsh environments and for guiding the development of more robust units.

## ASSOCIATED CONTENT

**Supporting Information**. M(bdc)(ted)$_{0.5}$ structure description, The IR adsorption of M(bdc)(ted)$_{0.5}$ [M=Zn, Ni and Co], Raman spectra of activated M(bdc)(ted)$_{0.5}$ [M=Zn, Ni and Co] and hydrated M(bdc)(ted)$_{0.5}$ , Raman spectra of M(bdc)(dmf)[M=Zn, Cu], Raman spectra of Cu(bdc)(ted)$_{0.5}$ during hydration under 6 Torr, X-ray diffraction pattern of Cu(bdc)(ted)$_{0.5}$ during hydration under 6 Torr, X-ray diffraction patterns of Cu(bdc)·(dmf), Zn(bdc)(dmf) and MOF-2 structure, cyclic adsorption/desorption of water experiment, IR spectra of D$_2$O adsorption into M(bdc)(ted)$_{0.5}$ [M=Zn, Ni, Co] and integrated areas of stretching band in the region of 2736-1753 cm$^{-1}$ as a function of vapor pressure.

## AUTHOR INFORMATION
### Corresponding Author
* Phone 1-972-883-5751; e-mail: chabal@utdallas.edu.

## ACKNOWLEDGMENT
This work was supported in its totality by the Department of Energy, Basic Energy Sciences, division of Materials Sciences and Engineering (DOE grant No. DE-FG02-08ER46491).

Graphic for Table of Contents

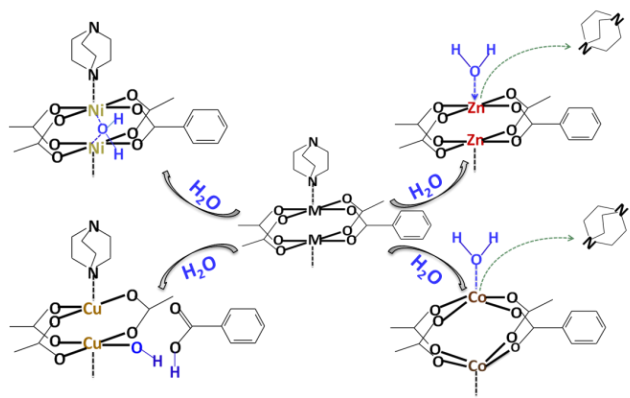